\documentstyle[twoside,fleqn,espcrc2,epsf]{article}

\newcommand{\AmS}{{\protect\the\textfont2
  A\kern-.1667em\lower.5ex\hbox{M}\kern-.125emS}}

\def\spose#1{\hbox to 0pt{#1\hss}}
\def\ltapprox{\mathrel{\spose{\lower 3pt\hbox{$\mathchar"218$}}
 \raise 2.0pt\hbox{$\mathchar"13C$}}}
\def\gtapprox{\mathrel{\spose{\lower 3pt\hbox{$\mathchar"218$}}
 \raise 2.0pt\hbox{$\mathchar"13E$}}}
\def\inapprox{\mathrel{\spose{\lower 3pt\hbox{$\mathchar"218$}}
 \raise 2.0pt\hbox{$\mathchar"232$}}}

\newcommand{\be}{\begin{equation}}
\newcommand{\ee}{\end{equation}}

\def\su3{$SU(3)$}

\newcommand{\sig}{{a^2 \sigma}}
\newcommand{\ssig}{{\sqrt{a^2 \sigma}}}

\newcommand{\mpi}{{a m_\pi}}

\newcommand{\mrho}{{a m_\rho}}
\newcommand{\mN}{{a m_N}}
\newcommand{\mDelta}{{a m_\Delta}}

\def\order{{\cal O}}

\title{Comparing Wilson and Clover quenched $SU(3)$ spectroscopy with
  an improved gauge action}

\author{
  Sara~Collins\address{Dept. of Physics, Glasgow, Scotland},
  Robert~G.~Edwards,${}^{\rm b,}$\thanks{Speaker at the conference.
      FSU-SCRI-96-70.
      This research was supported by DOE contracts
      DE-FG05-85ER250000 and DE-FG05-92ER40742.}, 
  Urs~M.~Heller${}^{\rm b}$, 
  and 
  John~Sloan\address{SCRI, Florida State University, 
      Tallahassee, FL 32306-4052, USA}
}
\begin{document}

\begin{abstract}

We present results of quenched $SU(3)$ hadron spectroscopy comparing
$\order(a)$ improved Wilson (Clover) fermions with conventional Wilson
fermions.  The configurations were generated using an $\order(a^2)$
improved 6-link $SU(3)$ pure gauge action at $\beta$'s corresponding
to lattice spacings of $0.15$, $0.18$, $0.20$, $0.33$, and $0.43$ fm. 
We find evidence that fermionic scaling violations are consistent with
$\order(a^2)$ for Clover and $\order(a)$ with a nonnegligible
$\order(a^2)$ term for standard Wilson fermions. This latter mixed
ansatz makes a reliable continuum extrapolation problematic for Wilson
fermions.  We also find that the slope of the scaling violations is
roughly $250 MeV$ for both Wilson and Clover fermions.

\end{abstract}

\maketitle

\section{INTRODUCTION}

The Symanzik action improvement program has been 
proposed\cite{Luscher_85,SW_85} as a way to reduce scaling violations 
in the approach to the continuum limit from a lattice action.  In this
contribution, we report on our investigations into the nature of
scaling violations inherent in improved actions.  We perform $SU(3)$
quenched spectroscopy using a one-loop tadpole-improved Symanzik pure
gauge action and compare the tree-level tadpole-improved Clover
fermion action with the standard Wilson fermionic action. We measure
the hadron spectrum and the string tension at lattice spacings $0.15$,
$0.18$, $0.20$, $0.25$, $0.33$, and $0.43$ fm. The goal of this work
is to measure the lattice spacing dependence of the hadron
spectrum. We find significant scaling violations in $\mrho/\ssig$
consistent with $\order(a^2)$ for Clover fermions and $\order(a)$ for
Wilson fermions. However, Wilson fermions show a nonnegligible
$\order(a^2)$ term. Without this correction, continuum
extrapolations of $\mrho/\ssig$ decrease with increasing
$\ssig$; hence, reliable continuum extrapolations are problematic.

\section{ACTIONS}

In this work, we used the tadpole improved 1-loop correct Symanzik
pure gauge action described in Ref.\cite{Luscher_85,Alford_95}.
Classically, this action has $\order(a^4)$ errors but quantum effects
induce $\order(\alpha^2 a^2)$ errors.  However, it was shown
\cite{Alford_95} that the 
gauge action is insensitive to nonperturbative tuning of the
coefficients suggesting that the $\order(\alpha^2 a^2)$ errors
are small. 	

\begin{figure}[t]
\vspace*{-10mm} \hspace*{-0cm}
\begin{center}
\epsfxsize = 0.45\textwidth
\leavevmode\epsffile{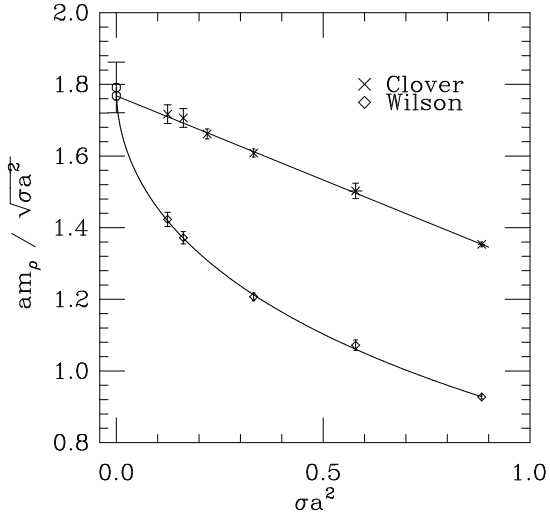}
\end{center}
\vspace*{-2cm}
\caption{
  Scaling plot of $\mrho$ versus the string tension $\sig$.
}
\label{rho_sig_fig}
\end{figure}

We compute hadron spectroscopy using the Wilson action, and the
tadpole-improved version of the tree-level Clover fermionic action
proposed in \cite{SW_85}. With this tree-level action, one expects
quantum errors of $\order(\alpha a)$. However, Naik \cite{Naik_93}
computed the one loop and the dominant two loop contributions to
$c(g_0^2)$ and finds that after tadpole improvement the coefficient of
$g_0^2$ is $0.016$.  Therefore, we expect to see only the dominant
$\order(a^2)$ fermionic errors.

Recently, the Clover coefficient with the Wilson gauge action was
computed nonpertubatively~\cite{Alpha_96} and was found to deviate
significantly from the tadpole value. Since our gauge 
actions differ from~\cite{Alpha_96} at $\order(a^2)$, we
cannot use this result in the present work.

\begin{figure}[t]
\vspace*{-10mm} \hspace*{-0cm}
\begin{center}
\epsfxsize = 0.45\textwidth
\leavevmode\epsffile{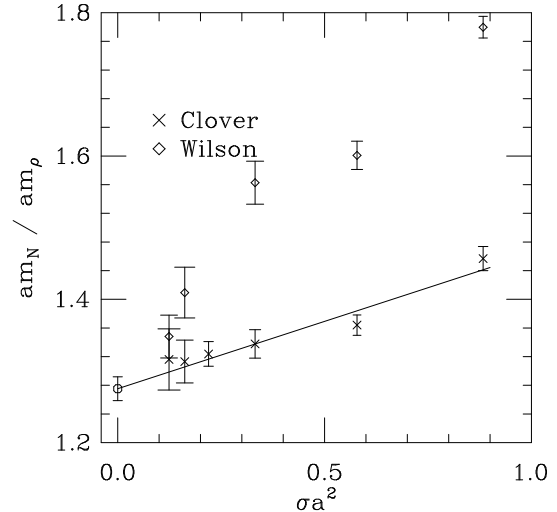}
\end{center}
\vspace*{-2cm}
\caption{
  Scaling plot of the nucleon versus the rho as a function of the
  string tension $\sig$. 
}
\label{N_rho_sig_fig}
\end{figure}

\section{OBSERVABLES}

To set the scale, $a$, we used the string tension.  We computed finite
$T$ approximations to the static quark potential using time-like
Wilson loops $W(\vec R, T)$ which were constructed using `APE'-smeared
spatial links \cite{APE_87,Collins_lat95}. We extracted the string
tension from the fitted ``effective'' potentials.


For hadron measurements, we used correlated multi-state fits to
multiple correlation functions as discussed in \cite{Collins_lat95}. We
computed quark propagators in Coulomb gauge at several $\kappa$ values
for each $\beta$. We used two gaussian source smearing functions with
smeared and local sinks.  
Fitting ranges were chosen by an automated procedure~\cite{Collins_lat95}.


\section{SIMULATIONS AND RESULTS}

We generated $100$ quenched configurations using the Symanzik gauge
action \cite{Luscher_85,Alford_95} on a $16^3\times32$ lattice at
$\beta = 7.90$, $7.75$, $7.60$, $7.40$, $7.10$, and $6.80$.  We expect
finite volume effects to be small since the box size is $2.40$fm at
$\beta=7.90$.  We originally had problems with exceptional
configurations on $8^3\times16$ at $\beta=6.80$. With the larger
physical volume, we reduced this problem. The cost of inversions
increased; however, we compensated for this by using the method of
`$Z(3)$' fermion sources~\cite{Butler_94} to increase the statistics
per inversion.

The results of the chiral
extrapolations of our best fits are listed in Table~\ref{fit_table}
along with the string tension. The lattice spacings we quote used
$\ssig = 440{\rm MeV}$.

\begin{table*}[t]
\begin{center}
\small
\caption{String tension and best fits of masses extrapolated to
physical ratios for Clover (top) and Wilson(bottom).}
\label{fit_table}
\begin{tabular}{|c|c|l|l|l|l|l|l|l|}
\hline
$\beta$ & $u_0$ & $\sig$ & $\mpi$ & $\mrho$ & $J$ & $\mN$ & $\mDelta$ 
& $\mrho/\ssig$ \\
\hline
$7.90$ & 0.8848 & 0.1240( 27) & 0.1084(12) & 0.6046(65) &
  0.3788(133) & 0.796(26) & 1.050(52) & 1.717(26) \\
       &        &             & 0.0898( 7) & 0.5012(36) & 
  0.3381( 32) & 0.656(15) & 0.844(16) & 1.423(20) \\
\hline
$7.75$ & 0.8800 & 0.1620( 32) & 0.1231(14) & 0.6868(76) &
  0.3863( 88) & 0.902(20) & 1.155(30) & 1.706(26) \\
       &        &             & 0.0990( 8) & 0.5522(42) & 
  0.3498( 55) & 0.778(19) & 0.948(52) & 1.372(17) \\
\hline
$7.60$ & 0.8736 & 0.2196( 26) & 0.1396( 8) & 0.7787(47) &
  0.3796( 47) & 1.031(14) & 1.338(30) & 1.662(14) \\
\hline
$7.40$ & 0.8629 & 0.3320( 26) & 0.1661( 9) & 0.9266(52) &
  0.3706( 48) & 1.240(18) & 1.560(31) & 1.608(11) \\
       &        &             & 0.1246(10) & 0.6954(55) & 
  0.3352( 52) & 1.087(20) & 1.308(18) & 1.207(11) \\
\hline
$7.10$ & 0.8441 & 0.5785(154) & 0.2048(10) & 1.1429(56) &
  0.3498( 38) & 1.559(16) & 2.006(34) & 1.503(22) \\
       &        &             & 0.1461( 7) & 0.8153(37) & 
  0.3178( 25) & 1.305(15) & 1.548(14) & 1.072(15) \\
\hline
$6.80$ & 0.8261 & 0.8831( 75) & 0.2279( 9) & 1.2719(52) &
  0.3448( 37) & 1.853(21) &           & 1.354( 7) \\
       &        &             & 0.1562( 7) & 0.8717(36) & 
  0.3140( 30) & 1.551(12) & 1.747(11) & 0.928( 6) \\
\hline
\end{tabular}
\end{center}
\end{table*}

Our main result for the determination of scaling violations is in
Figure~\ref{rho_sig_fig}. We extrapolate the vector mass to the
physical $\rho/\pi$ ratio and plot the result of $\mrho/\ssig$ for
Wilson and Clover. Fitting the functional form of the scaling
violations, we find that the confidence level $Q$ is $0.37$ for Wilson
assuming $\order(a)$ errors and $10^{-6}$ assuming $\order(a^2)$. We
are confident then that the Wilson scaling violations are not
consistent with only $\order(a^2)$. However, as we drop the lower
$\beta$ points in the $\order(a)$ ansatz, we $Q$ and the
intercept rise indicating that quadratic scaling corrections are not
small. This is not too surprising since the Clover result varies by
$23\%$ over the range of the fit while the Wilson result varies $48\%$.
Fitting to both $\order(a)$ and $\order(a^2)$ we have $Q=0.62$ and an
intercept of $1.79(7)$ with an $\order(a^2)$ term that is two standard
deviations away from $0$. This mixed ansatz is plotted in
Figure~\ref{rho_sig_fig}.

For Clover, unfortunately, $Q$ is large for both $\order(a)$ and
$\order(a^2)$ ans\"atze; hence, we can not discern the scaling
violation order solely by $Q$. However, the extrapolated value of
$\mrho/\ssig$ is $1.98(2)$ compared to $1.77(1)$ for $\order(a)$ and
$\order(a^2)$ ans\"atze, resp. Hence, only the $\order(a^2)$ ansatz is
compatible with the Wilson continuum prediction.  We find that the
$\order(a^2)$ fit is stable after dropping low $\beta$ points.
Furthermore, the GF11 and MILC collaborations find a consistent
continuum extrapolation using Wilson and Staggered fermions, resp.
Using the mixed $\order(a)$ and $\order(a^2)$ ansatz for Clover, we
find the fit has a linear term that is zero within errors; hence, we will
ignore this fit. To adequately resolve the problem of $\order(a)$ and
$\order(a^2)$ scaling, we would need a simulation at $a \ltapprox
0.08$fm.

Another main feature of the extrapolation is the slope. For Wilson
fermions, we find the coefficients of the $\order(a)$ and
$\order(a^2)$ terms (the slopes) are $280$ MeV and $(160 {\rm MeV})^2$. 
For Clover, we find the coefficient of the $\order(a^2)$
term is $(230 {\rm MeV})^2$. These slopes are consistent with the
characteristic soft scale of light spectroscopy 
$\Lambda_{\rm QCD}$. We find smaller slopes for the nucleon and
delta.

In Figure~\ref{N_rho_sig_fig}, we plot the ratio of $\mN / \mrho$ assuming
$\order(a^2)$ scaling violations for Clover. We find a continuum value
of $1.28(2)$ that is consistent with the GF11 result\cite{Butler_94};
however, it is different than the experimental value of $1.22$. For
the delta, we obtain a continuum value of $1.67(4)$ for Clover
compared to $1.61(8)$ for GF11. 
Our Wilson continuum extrapolations using only an $\order(a)$ ansatz are
consistently below the Clover and GF11 extrapolations.


%
%
%

\end{document}